\pdfoutput=1

\documentclass{raa}           

\usepackage{graphicx,times}
\usepackage{natbib}
\usepackage{amssymb,amsmath}
\bibpunct{(}{)}{;}{a}{}{,}

\usepackage[pagebackref=true]{hyperref}

\begin{document}

   \title{Power-law distribution and scale-invariant structure from the first CHIME/FRB Fast Radio Burst catalog}

 \volnopage{ {\bf 20XX} Vol.\ {\bf X} No. {\bf XX}, 000--000}
   \setcounter{page}{1}

   \author{Zi-Han Wang \inst{1}, Yu Sang\inst{1,2}\footnote{Corresponding author}, Xue Zhang\inst{1,2}
   }

   \institute{ Center for Gravitation and Cosmology, College of Physical Science and Technology, Yangzhou University, Yangzhou 225009, China; {\it sangyu@yzu.edu.cn}\\
        \and
             Shanghai Frontier Science Research Center for Gravitational Wave Detection, School of Aeronautics and Astronautics, Shanghai Jiao Tong University, Shanghai 200240, China\\
\vs \no
   {\small Received 20XX Month Day; accepted 20XX Month Day}
}

\abstract{We study the statistical property of fast radio bursts (FRBs) based on a selected sample of 190 one-off FRBs in the first CHIME/FRB catalog. Three power law models are used in the analysis, and we find the cumulative distribution functions of energy can be well fitted by bent power law and thresholded power law models. And the distribution functions of fluctuations of energy well follow the Tsallis $q$-Gaussian distribution. The $q$ values in the Tsallis $q$-Gaussian distribution are constant with small fluctuations for different temporal scale intervals, indicating a scale-invariant structure of the bursts. The earthquakes and soft gamma repeaters show similar properties, which are consistent with the predictions of self-organized criticality systems.
}

   \authorrunning{Z.-H. Wang et al. }            
   \titlerunning{Power-law distribution of FRBs}  
   \maketitle

%
\section{Introduction}           
\label{sec:introduction}
Fast radio bursts (FRBs) are highly energetic transients with a few milliseconds durations occurring at cosmological distances \citep{Petroff:2019tty,Cordes:2019cmq,Zhang:2020qgp,Xiao:2021omr}. 
In general FRBs are observationally classified into repeaters and non-repeaters, where the repeating FRBs are multiple bursts detected from the same source and non-repeating FRB is a one-off burst.
The nature of FRBs remains a mystery, although hundreds of FRBs have been detected since the the discovery of the first event \citep{Lorimer:2007qn}.
Many models have been proposed to explain the origin of FRBs \citep{Platts:2018hiy}, but none of them have been fully confirmed.

The distribution of FRBs energy has been found to follow a power law function \citep{Lu:2016fgg,Li:2016qbl,Wang:2016lhy,Macquart:2018jlq,Wang:2017agh,Lu:2019pdn,Wang:2019sio,Wang:2019suh,Lin:2019ldn}. 
\citet{Wang:2016lhy} analysed 17 bursts from repeating FRB 121102 and found the cumulative distributions of duration, peak flux and fluence follow the power-law forms, which is similar to the statistical properties of soft gamma repeaters (SGRs).
\citet{Wang:2017agh} found the 14 new bursts of FRB 121102 detected by the Green Bank Telescope have a power-law energy distribution, very similar to the Gutenberg-Richter law found in earthquakes. 
20 FRBs of the Australian Square Kilometre Array Pathfinder (ASKAP) sample including repeaters and non-repeaters were also found to be well described with a power-law energy distribution \citep{Lu:2019pdn}.
\citet{Wang:2019sio} used six samples of the repeating FRB 121102 observed by different telescopes at different frequencies and found energy has a universal power-law distribution and the power-law index is similar to nonrepeating FRBs.
Using two samples  of the repeating FRB 121102 from  Green Bank Telescope and Arecibo Observatory, the cumulative distributions of flux, fluence, energy and waiting time have been found to follow the bent power law \citep{Lin:2019ldn}.

The power law distribution is a very common statistical property of a lot of astrophysical phenomena. The power law size distributions of waiting time, duration, flux and fluence are predictions of self-organized criticality (SOC) system  \citep{aschwanden2011self,Bak:1987xua,Wang:2013wha}, which exist in a large number of natural systems exhibiting non-linear energy dissipation.
For example, earthquakes, solar flares and SGRs have been explained by an SOC system \citep{Bak:1989Earth,Lu:1991ApJ,Cheng1996}.  \citet{Cheng1996} compared the the statistical properties of 111 bursts from SGR 1806-20  observed during 1979-1984 with earthquakes and found  the cumulative distribution of energy in SGRs is similar to the well-known Gutenberg-Richter power law in earthquakes with near power law index, as well as the similarity of waiting time and duration distributions between SGRs and earthquakes.

Another predicted behaviour of  SOC system is the scale invariance structure of the energy fluctuations, which have been studied in earthquakes \citep{Caruso:2007PhRvE,Wang:2015nsl}, SGRs \citep{Chang:2017bnb,Wei:2021kdw,Sang:2021cjq} and FRBs \citep{Lin:2019ldn,Wei:2021kdw}.
The probability density functions (PDFs) of fluctuations in earthquake energy follow a $q$-Gaussian form with fat tails \citep{Caruso:2007PhRvE}. 
The PDFs of energy fluctuations do not depend on the scale intervals, namely, the $q$-Gaussian distributions parameter $q$ in  are approximately equal for different scale intervals, indicating a scale-invariant structure in the energy fluctuations of earthquakes  \citep{Wang:2015nsl}. 
The PDFs of energy fluctuations at different scale intervals of SGRs and repeating FRBs have also exhibit a universal $q$-Gaussian distribution, implying the underlying association of the origins between SGRs and repeating FRBs \citep{Chang:2017bnb,Lin:2019ldn,Wei:2021kdw,Sang:2021cjq}.

Recently the Canadian Hydrogen Intensity Mapping Experiment Fast Radio Burst (CHIME/FRB) project released the first catalog of FRBs detected from 2018 July 25 to 2019 July 1 \citep{CHIMEFRB:2021srp}.
In this paper, we use 190 one-off FRBs in the CHIME/FRB catalog to study the statistical properties of energy. In Section \ref{sec:CDF}, we study the cumulative distribution of energy. In Section \ref{sec:fluctuations}, we study the cumulative distribution of fluctuations. Finally, discussion and conclusion are given in Section \ref{sec:conclusions}.

\section{the cumulative distribution function of energy}\label{sec:CDF}
The first CHIME/FRB catalog consists of 536 FRBs detected between 400 and 800 MHz from 2018 July 25 to 2019 July 1, including 62 bursts from repeaters and  474 one-off bursts \citep{CHIMEFRB:2021srp}. We use a selected sample of 190 one-off FRBs in the CHIME/FRB catalog to study the statistical properties of energy. Given that bursts with low fluences are missed by the CHIME detection algorithm \citep{CHIMEFRB:2021srp,Hashimoto:2022llm}, we select reliable sample with a certain fluence threshold ($> 5~ {\rm Jy~ms}$) as a correction to observational incompleteness. The measured fluence and flux are dependent on the distance of each source, so we study the statistical properties of energy instead. We use the rest-frame isotropic radio energy $E_{{\rm rest},400}$, the calculation detail of which can be found in \citet{Hashimoto:2022llm}.

\begin{figure}
 \centering
 \includegraphics[width=0.8\textwidth]{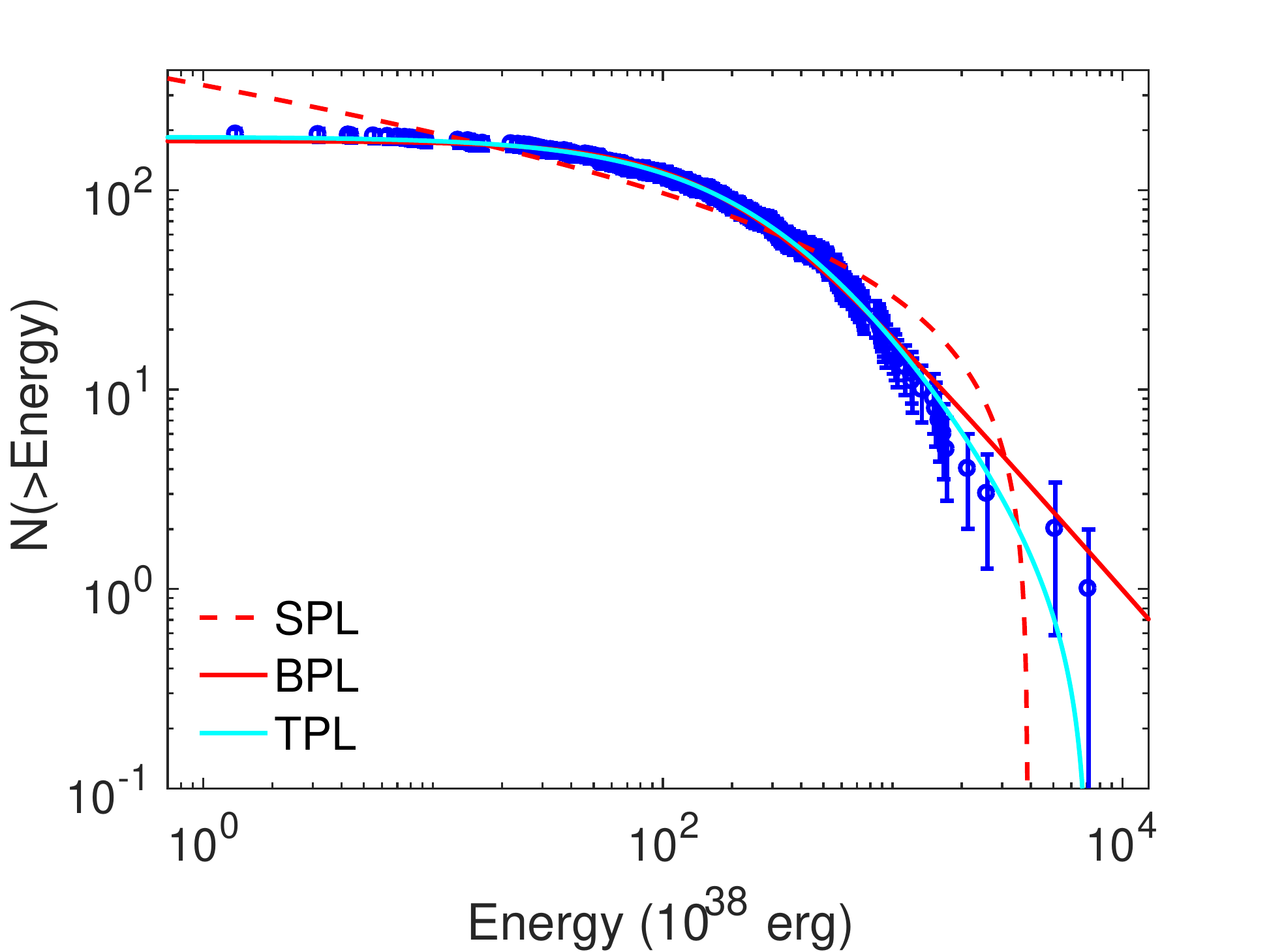}
 \caption{The CDFs of energy. The red dashed, red solid, and cyan solid lines are the best-fitting curves to the SPL, BPL, and TPL models, respectively. }
 \label{fig:cdf}
\end{figure}

The cumulative distribution functions (CDFs) of energy are shown by blue errorbars in Figure \ref{fig:cdf}.
As discussion in Section \ref{sec:introduction}, power law distribution of energy is predicted by SOC theory. Here we use  the simple power law (SPL) model, the bent power law (BPL) model and the thresholded power law (TPL) model to fit the data.
The best-fitting parameters are calculated by minimizing the $\chi^2$ statistics,
\begin{equation}
  \chi^2=\sum_i\frac{[N_i-N(>x_i)]^2}{\sigma_i^2},
\end{equation}
where we take the uncertainty of data point as $\sigma_i=\sqrt{N_i}$.

The first model we use is the SPL model, which has been used to fit the CDFs of duration, peak flux and fluence of the repeating FRB 121102 \citep{Wang:2016lhy}.
The SPL model is given by
\begin{equation}
  N(>x) = A (x^{-\alpha}-x_c^{-\alpha}), ~~~ x<x_c,
\end{equation}
where $A$ is a normalization factor, $\alpha$ is the power-law index and $x_c$ is the cut-off value above which $N(>x_c)=0$. 
We list the best-fitting parameters in Table \ref{tab:para_cdf}. For the SPL model, the power law index $\alpha$ of energy is $0.16\pm0.02$. The cut-off parameter $x_c$ of energy is $(3880.41\pm255.37)\times10^{38}~{\rm erg}$. The best-fitting curves to the SPL model are shown in the red dashed lines in Figure \ref{fig:cdf}. The SPL models are consistent with the observation in the middle part but fit the data poorly at the left and right ends.

The second model we use is the BPL model. It has been used to fit the energy power density spectrum of gamma-ray bursts \citep{Guidorzi:2016ddt}, as well as the CDFs of energy of SGRs \citep{Chang:2017bnb,Sang:2021cjq} and FRBs \citep{Lin:2019ldn}.
The BPL model is given by
\begin{equation}
  N(>x)=B\left[1+\left(\frac{x}{x_b}\right)^{\beta}\right]^{-1},
\end{equation}
where $B$ is a parameter denoting the total number of the bursts, $x_b$ is the median value of $x$, i.e. $N(x>x_b)=B/2$, and $\beta$ is the power law index. 
The BPL is a smoothly connected piecewise function with a break at around $x_b$, below which the function is a plateau with power law index $p = 0$, and above which the function is reduced to be a simple power law with power law index $p = \beta$.
As a phenomenological model, the underlying physical meaning of BPL model is unknown currently.
For the BPL model, the power law index $\beta$ of energy is $1.31\pm0.02$. The median values $x_b$ of energy is $(191.54\pm3.51)\times10^{38}~{\rm erg}$. The best-fitting curves to the BPL model are shown in the red solid lines in Figure \ref{fig:cdf}. The BPL models are consistent with the observation very well for most of the data range. The best fitting BPL function is out of the 1 $\sigma$ confidence interval of data at around $2\times10^{41}~{\rm erg}$. 
 
The third model we use is the TPL model, which has been used to fit the energy distributions of magnetar bursts and  FRBs \citep{Cheng:2019ykn}.
Integrating the differential distribution of the TPL model, 
\begin{equation}
   N(x)dx = n_0 (x_0 + x)^{- \gamma } dx,\quad x_1\leq x\leq x_2,
\end{equation}
we can get the cumulative distribution as 
\begin{equation}
   N( >x) =  \int_{x}^{x_2} N(x)dx  = \frac{n_0}{1-  \gamma } [(x_0 + x_2)^{1- \gamma } - (x_0 + x)^{1- \gamma } ].
\end{equation}
Here $x_0$ is a constant added to the ideal power law distribution function, $n_0$ is a normalization constant and $\gamma$ is the power law index. $x_1$ and $x_2$ are the lower and upper cutoffs of the data points, respectively. The free parameters in the TPL model are $(n_0, x_0, \gamma)$.
For the TPL model, the power law index $\gamma$ of energy is $2.79\pm0.05$. The parameter $x_0$ of energy is   
$(381.79\pm16.89)\times10^{38}~{\rm erg}$. The best-fitting curves to the TPL model are shown in the cyan solid lines in Figure \ref{fig:cdf}. The TPL models are consistent with the observation very well. We list the $\chi^2_{\rm red}$ values for the SPL, BPL and TPL models in Table \ref{tab:para_cdf}. Throughout the three models, the TPL model is most consistent with observation.

\begin{table}
  \centering
  \caption{The best-fitting parameters to the SPL, BPL and TPL models. The units of $x_c$ ($x_b$ and $x_0$) for energy are $10^{38}~{\rm erg}$. The reduced chi-square  $\chi^2_{\rm red}$ is the chi-square values per degree of freedom.}
  \label{tab:para_cdf}
  \begin{tabular}{cccccc}
  \hline
  Model & Parameters    \\
  \hline
  & $\alpha$ = $0.16\pm0.02$  \\
  SPL & $x_c$ = $3880.41\pm255.37$  \\
  & $\chi^2_{\rm red}$  = $5.20$  \\
  \hline
  & $\beta$ = $1.31\pm0.02$   \\
  BPL & $x_b$ = $191.54\pm3.51$   \\
  & $\chi^2_{\rm red}$ = $0.29$  \\
  \hline
  & $\gamma$ = $2.79\pm0.05$  \\
  TPL & $x_0$ = $381.79\pm16.89$  \\
  & $\chi^2_{\rm red}$ = $0.12$  \\
  \hline
  \end{tabular}
\end{table}

\section{Cumulative distribution functions of fluctuations}\label{sec:fluctuations}
In this section, we study the statistical property of fluctuations of energy. 
The fluctuation is defined as $Z_n=Q_{i+n}-Q_i$, and here $Q_i$ is the observed quantity from the $i$-th burst in temporal order. The integer $n$ is the temporal interval scale of interest. We use the dimensionless fluctuation rescaled by the standard deviation $z_n=Z_n/\sigma$, where $\sigma={\rm std}(Z_n)$. In our case, the $Q$ quantity refers to energy.
The fluctuations from earthquakes, SGRs and FRBs have been shown to follow  the Tsallis $q$-Gaussian distribution \citep{Wang:2015nsl,Chang:2017bnb,Wei:2021kdw,Sang:2021cjq,Lin:2019ldn}. 

The $q$-Gaussian function is given by \citep{Tsallis:1987eu,Tsallis:1998ws}
\begin{equation}
  f(x) = a [1 - b (1-q) x^2]^{\frac{1}{1-q}}.
\end{equation} 
Here $a$ is the normalization factor, and the parameters $b$ and $q$ are corresponding to the width and sharpness of the peak, respectively. As a generalization of the Gaussian distribution, the $q$-Gaussian function has a sharper peak and fatter tails. 
The parameter $q$ describes the deviation from the Gaussian distribution, for example, in the limit  $q\rightarrow 1$, the $q$-Gaussian function will reduce to the Gaussian function with mean $\mu = 0$ and standard deviation $\sigma=1/\sqrt{2 b}$. 
In practice we use the CDFs of Tsallis $q$-Gaussian function to fit the fluctuations data, which is calculated by 
\begin{equation}
  F(x)=\int_{-\infty}^xf(x)dx.
\end{equation}
We take the fluctuations in temporal interval scale $n$ from 1 to 40.

\begin{figure}
 \centering
 \includegraphics[width=0.8\textwidth]{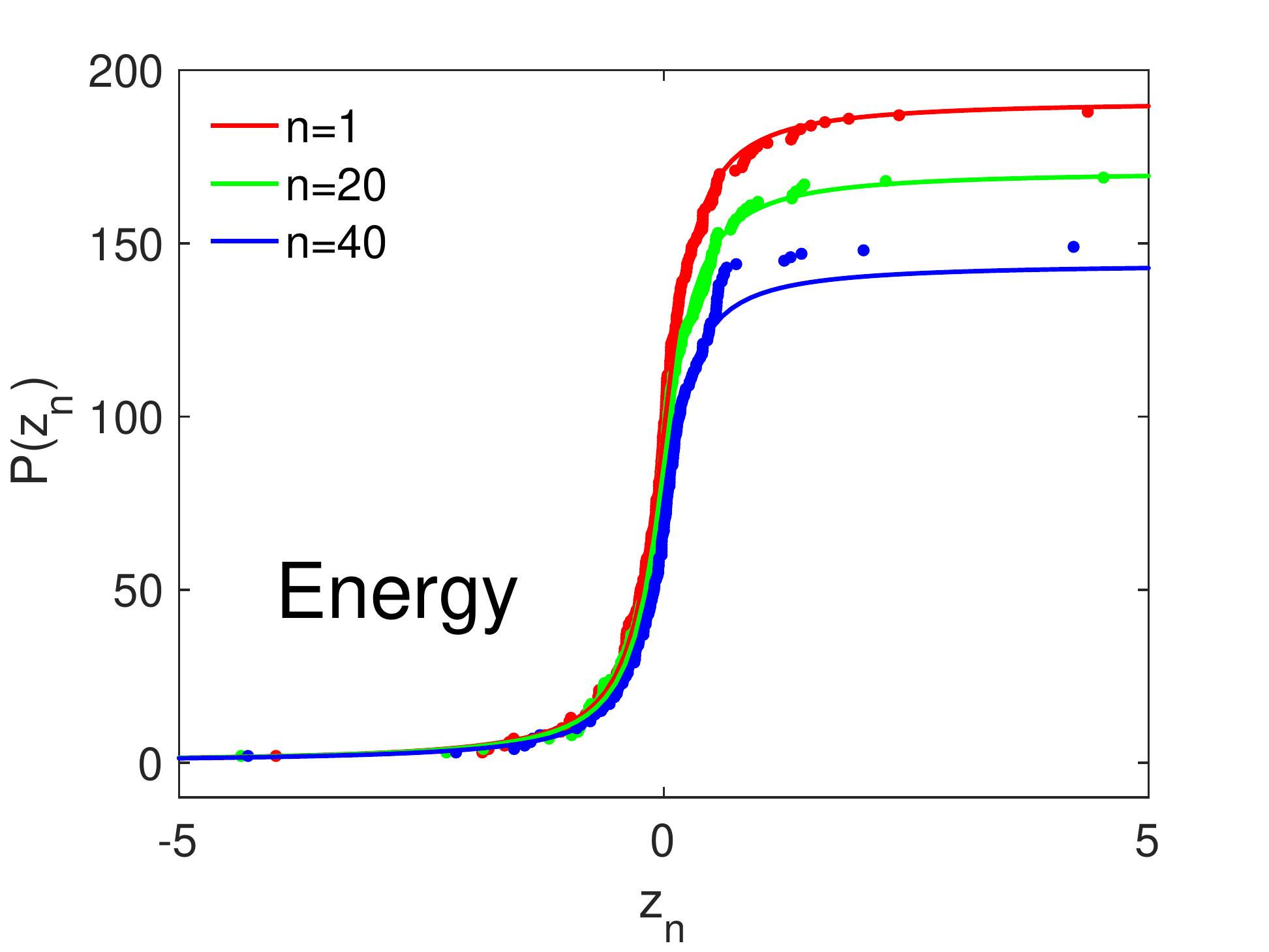}
 \caption{Examples of the CDFs of fluctuation of energy.  }\label{fig:fluctuation}
\end{figure}

The CDFs of fluctuations with interval scale $n = 1,20,40$ of energy are shown in Figure \ref{fig:fluctuation}.
The best-fitting curves to the CDF of $q$-Gaussian function are shown in red, green and blue solid lines with temporal interval scale $n = 1,20,40$, respectively.
The best-fitting $q$ values for $n = 1,20,40$ are listed in Table \ref{tab:para_q}.
The Tsallis $q$-Gaussian model is consistent with the observation data very well, except for the fitting of right part with $n = 40$. The poor fitting is because of the less data points for fluctuations at $z_n > 1$ with temporal interval scale $n = 40$. As shown by the blue points in Figure  \ref{fig:fluctuation}, only 5 energy fluctuations at $z_n > 1$ are used in the fitting.

\begin{table}
  \centering
  \caption{The best-fitting $q$ values for $n = 1,20,40$.}
  \label{tab:para_q}
  \begin{tabular}{cccccc}
  \hline
   & $q$ values   \\
  \hline
   $n=1$ & $1.87\pm0.01$  \\
  $n=20$ & $1.89\pm0.02$  \\
   $n=40$  & $1.91\pm0.02$  \\
  \hline
  \end{tabular}
\end{table}

\begin{figure}
 \centering
 \includegraphics[width=0.8\textwidth]{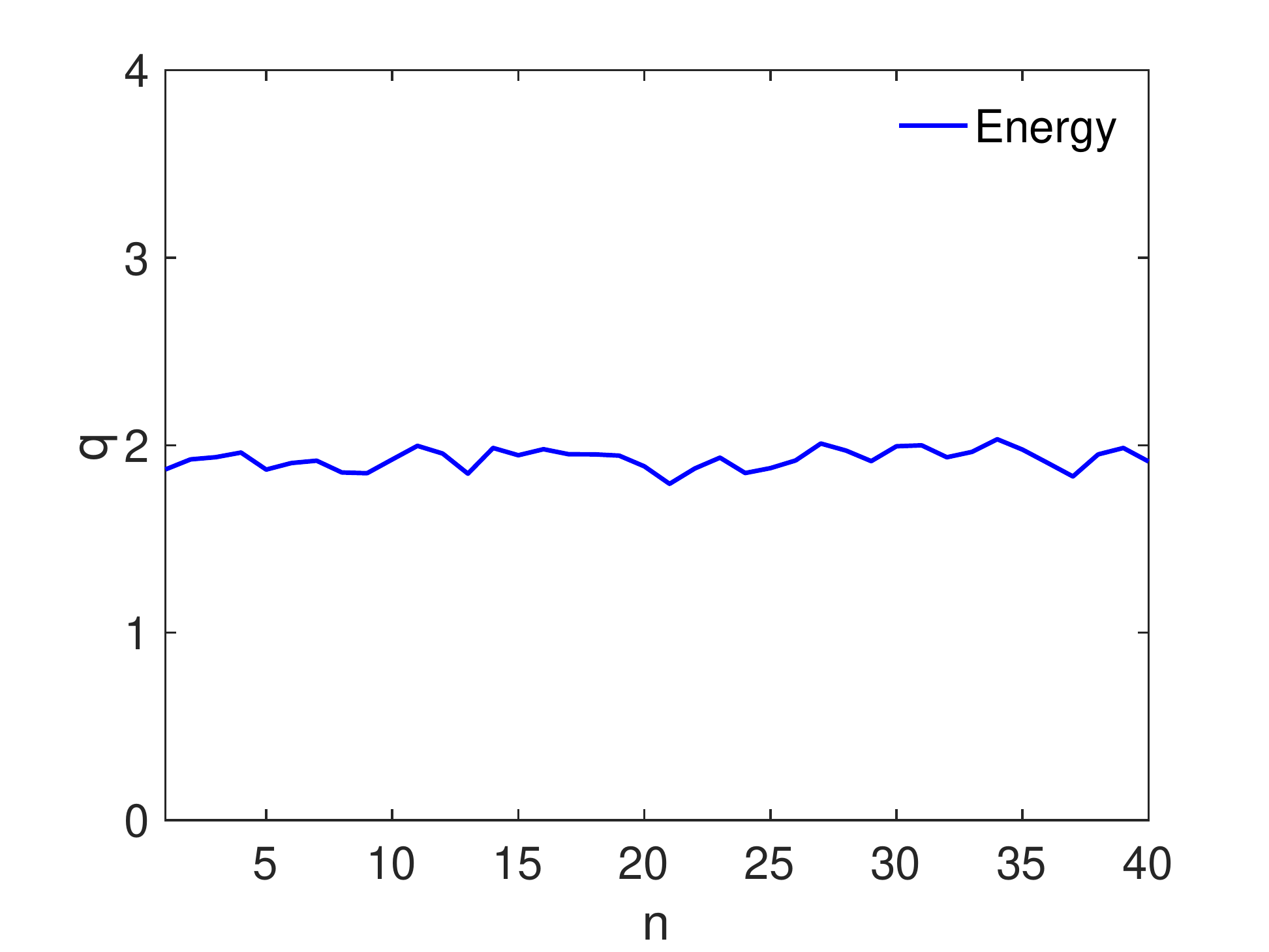}
 \caption{The best-fitting $q$ values as a function  $n$. }\label{fig:q}
\end{figure}

We further study the scale invariance property in the temporal interval $n$ of the bursts.
We fit the CDFs of fluctuations of energy for all the temporal interval scale $n $ from 1 to 40.
The best-fitting parameter $q$ as a function of the temporal interval scale $n$ is shown in Figure \ref{fig:q}.
The $q$ value is a constant with small fluctuations for energy. The mean values and standard deviations of $q$ for $n = 1 - 40$ for energy are $1.93\pm0.05$.
The independence on interval scale $n$ of the fluctuations of energy indicates a scale invariance structure in the property of bursts.

\section{Discussion and Conclusions}\label{sec:conclusions}
In summary, we have investigated the statistical properties of 190 one-off FRBs in the first CHIME/FRB catalog.  The CDFs of energy are fitted using three power law models, and we found the TPL model is most consistent with observation. We also studied the statistical properties of the fluctuations of energy, and found the distribution function of fluctuations well follows the Tsallis $q$-Gaussian distribution. More importantly, the $q$ values in the Tsallis $q$-Gaussian distribution are constant with small fluctuations for  different temporal scale intervals, indicating a scale-invariant structure of the bursts. The power law distribution of the energy and a scale-invariant Tsallis $q$-Gaussian distribution of fluctuations were found in the studies on earthquakes, SGRs and repeating FRBs. In this paper, we extend research on those features to non-repeating FRBs, supporting that the occurrence mechanism of FRBs might be explained by self-organized criticality model. The underlying assumption of our analysis is that all the one-off FRBs are following the same occurrence mechanism, which is the physical association of different sources.  In this sense, the FRBs from different sources could be thought as the repeating samples of the same process, for example, the SOC process. Another possibility is that most of these observed one-off FRBs are intrinsically repeaters. But only one burst was detected due to the limit of observational time and CHIME sensitivity. Although these bursts are detected as one-off FRBs, some may be reactive again in future.

Our analysis is based on the calculation of energy in \citet{Hashimoto:2022llm}. The dispersion measure of FRB host galaxy ${\rm DM}_{\rm host}$ is modelled by 50/(1+ $z$) ${\rm pc~cm}^{-3} $ in the derivation of the redshift for FRBs using the observed dispersion measure. The assumption of ${\rm DM}_{\rm host}$ is inconsistent with observations to some degree. For example, the ${\rm DM}_{\rm host}$ of FRB 20190520B is $\sim$ 900 ${\rm pc~cm}^{-3} $, nearly ten times higher than the average of FRB host galaxies \citep{Niu:2021bnl}, which may be contributed by the supernova remnant around the source \citep{Zhao:2021vns}. From numerical simulations, a reasonable distribution of ${\rm DM}_{\rm host}$ follows log-normal function \citep{Macquart:2020lln,Wu:2021jyk}. The model of ${\rm DM}_{\rm host}$  has an effect on the calculation of redshift and energy, and will influence our results of the energy fitting. For example, if ${\rm DM}_{\rm host}$ was underestimated, a larger $z$ would be derived, which will finally induce a larger energy.

\normalem
\begin{acknowledgements}
We would like to thank Hai-Nan Lin for helpful discussion. 
This work has been supported by the National Natural Science Foundation of China under Grant Nos. 12005184 and 12005183,
and the Natural Science Foundation of the Jiangsu Higher Education Institutions of China under Grant No. 20KJD140002.

\end{acknowledgements}
  
\bibliographystyle{raa}
\bibliography{reference}

\begin{thebibliography}{36}
\providecommand\natexlab[1]{#1}
\providecommand\JournalTitle[1]{#1}

\bibitem[Amiri {et~al.}(2021)]{CHIMEFRB:2021srp}
Amiri, M., {et~al.} 2021, Astrophys. J. Supp., 257, 59

\bibitem[Aschwanden(2011)]{aschwanden2011self}
Aschwanden, M. 2011, Self-Organized Criticality in Astrophysics: The Statistics
  of Nonlinear Processes in the Universe, Springer Praxis Books (Springer
  Berlin Heidelberg)

\bibitem[Bak \& Tang(1989)]{Bak:1989Earth}
Bak, P., \& Tang, C. 1989, Journal of Geophysical Research: Solid Earth, 94,
  15635

\bibitem[Bak {et~al.}(1987)]{Bak:1987xua}
Bak, P., Tang, C., \& Wiesenfeld, K. 1987, Phys. Rev. Lett., 59, 381

\bibitem[{Caruso} {et~al.}(2007)]{Caruso:2007PhRvE}
{Caruso}, F., {Pluchino}, A., {Latora}, V., {Vinciguerra}, S., \& {Rapisarda},
  A. 2007, Phys. Rev. E, 75, 055101

\bibitem[Chang {et~al.}(2017)]{Chang:2017bnb}
Chang, Z., Lin, H.-N., Sang, Y., \& Wang, P. 2017, Chin. Phys. C, 41, 065104

\bibitem[Cheng {et~al.}(1996)]{Cheng1996}
Cheng, B., Epstein, R.~I., Guyer, R.~A., \& Young, A.~C. 1996, Nature, 382, 518

\bibitem[Cheng {et~al.}(2020)]{Cheng:2019ykn}
Cheng, Y., Zhang, G.~Q., \& Wang, F.~Y. 2020, Mon. Not. Roy. Astron. Soc., 491,
  1498

\bibitem[Cordes \& Chatterjee(2019)]{Cordes:2019cmq}
Cordes, J.~M., \& Chatterjee, S. 2019, Ann. Rev. Astron. Astrophys., 57, 417

\bibitem[Guidorzi {et~al.}(2016)]{Guidorzi:2016ddt}
Guidorzi, C., Dichiara, S., \& Amati, L. 2016, Astron. Astrophys., 589, A98

\bibitem[Hashimoto {et~al.}(2022)]{Hashimoto:2022llm}
Hashimoto, T., {et~al.} 2022, Mon. Not. Roy. Astron. Soc., 511, 1961

\bibitem[Li {et~al.}(2017)]{Li:2016qbl}
Li, L., Huang, Y., Zhang, Z., Li, D., \& Li, B. 2017, Res. Astron. Astrophys.,
  17, 6

\bibitem[Lin \& Sang(2020)]{Lin:2019ldn}
Lin, H.-N., \& Sang, Y. 2020, Mon. Not. Roy. Astron. Soc., 491, 2156

\bibitem[Lorimer {et~al.}(2007)]{Lorimer:2007qn}
Lorimer, D.~R., Bailes, M., McLaughlin, M.~A., Narkevic, D.~J., \& Crawford, F.
  2007, Science, 318, 777

\bibitem[{Lu} \& {Hamilton}(1991)]{Lu:1991ApJ}
{Lu}, E.~T., \& {Hamilton}, R.~J. 1991, Astrophys. J., 380, L89

\bibitem[Lu \& Kumar(2016)]{Lu:2016fgg}
Lu, W., \& Kumar, P. 2016, Mon. Not. Roy. Astron. Soc., 461, L122

\bibitem[Lu \& Piro(2019)]{Lu:2019pdn}
Lu, W., \& Piro, A.~L. 2019, Astrophys. J., 883, 40

\bibitem[Macquart \& Ekers(2018)]{Macquart:2018jlq}
Macquart, J.-P., \& Ekers, R. 2018, Mon. Not. Roy. Astron. Soc., 480, 4211

\bibitem[Macquart {et~al.}(2020)]{Macquart:2020lln}
Macquart, J.~P., {et~al.} 2020, Nature, 581, 391

\bibitem[Niu {et~al.}(2022)]{Niu:2021bnl}
Niu, C.~H., {et~al.} 2022, Nature, 606, 873

\bibitem[Petroff {et~al.}(2019)]{Petroff:2019tty}
Petroff, E., Hessels, J. W.~T., \& Lorimer, D.~R. 2019, Astron. Astrophys.
  Rev., 27, 4

\bibitem[Platts {et~al.}(2019)]{Platts:2018hiy}
Platts, E., Weltman, A., Walters, A., {et~al.} 2019, Phys. Rept., 821, 1

\bibitem[Sang \& Lin(2022)]{Sang:2021cjq}
Sang, Y., \& Lin, H.-N. 2022, Mon. Not. Roy. Astron. Soc., 510, 1801

\bibitem[Tsallis(1988)]{Tsallis:1987eu}
Tsallis, C. 1988, J. Statist. Phys., 52, 479

\bibitem[Tsallis {et~al.}(1998)]{Tsallis:1998ws}
Tsallis, C., Mendes, R.~S., \& Plastino, A.~R. 1998, Physica A, 261, 534

\bibitem[Wang \& Dai(2013)]{Wang:2013wha}
Wang, F.~Y., \& Dai, Z.~G. 2013, Nature Physics, 9, 465

\bibitem[Wang \& Yu(2017)]{Wang:2016lhy}
Wang, F.~Y., \& Yu, H. 2017, JCAP, 03, 023

\bibitem[{Wang} \& {Zhang}(2019)]{Wang:2019sio}
{Wang}, F.~Y., \& {Zhang}, G.~Q. 2019, Astrophys. J., 882, 108

\bibitem[Wang {et~al.}(2021)]{Wang:2019suh}
Wang, F.~Y., Zhang, G.~Q., \& Dai, Z.~G. 2021, Mon. Not. Roy. Astron. Soc.,
  501, 3155

\bibitem[Wang {et~al.}(2015)]{Wang:2015nsl}
Wang, P., Chang, Z., Wang, H., \& Lu, H. 2015, Eur. Phys. J. B, 88, 206

\bibitem[Wang {et~al.}(2018)]{Wang:2017agh}
Wang, W., Luo, R., Yue, H., {et~al.} 2018, Astrophys. J., 852, 140

\bibitem[Wei {et~al.}(2021)]{Wei:2021kdw}
Wei, J.-J., Wu, X.-F., Dai, Z.-G., {et~al.} 2021, Astrophys. J., 920, 153

\bibitem[Wu {et~al.}(2022)]{Wu:2021jyk}
Wu, Q., Zhang, G.-Q., \& Wang, F.-Y. 2022, Mon. Not. Roy. Astron. Soc., 515, L1

\bibitem[Xiao {et~al.}(2021)]{Xiao:2021omr}
Xiao, D., Wang, F.~Y., \& Dai, Z.~G. 2021, Sci. China Phys. Mech. Astron., 64,
  249501

\bibitem[Zhang(2020)]{Zhang:2020qgp}
Zhang, B. 2020, Nature, 587, 45

\bibitem[Zhao \& Wang(2021)]{Zhao:2021vns}
Zhao, Z.-Y., \& Wang, F.~Y. 2021, Astrophys. J. Lett., 923, L17

\end{thebibliography}

\end{document}